\documentclass[aps,showpacs,showkeys,amsmath,amssymb,prl,reprint,superscriptaddress,lengthcheck]{revtex4-1}

\usepackage{graphicx}
\usepackage{dcolumn}
\usepackage{bm}
\usepackage{color}
\usepackage{amsmath}
\usepackage{amssymb}
\usepackage{gensymb}
\usepackage{slashed}
\usepackage{soul}
\usepackage{hyperref}
\hypersetup{colorlinks=true,
            linkcolor=blue,
            urlcolor=blue,
            citecolor=blue}

\definecolor{aogreen}{rgb}{0.0, 0.5, 0.0}
\definecolor{RiaanGreen}{RGB}{45,119,0}

\definecolor{revisedcolor}{RGB}{0,100,20}

\definecolor{Bcolor}{RGB}{10,200,10}
\definecolor{Jcolor}{RGB}{20,20,200}
\definecolor{Ccolor}{RGB}{200,20,20}
\definecolor{Qcolor}{RGB}{50,200,200}

\begin{document}
@book{Budker:2013,
  title={Optical Magnetometry},
  author={edited by D. Budker and D. F. Jackson Kimball},
  year={2013},
  publisher={Cambridge University Press, Cambridge, England}
}

@article{Budker/N:2007,
  title={Optical magnetometry},
  author={Budker, Dmitry and Romalis, Michael},
  journal={Nat. Phys.},
  volume={3},
  pages={227--234},
  year={2007},
  doi={10.1038/nphys566}
}

@article{Shah/NP:2007,
  title={Subpicotesla atomic magnetometry with a microfabricated vapour cell},
  author={Shah, V. and Knappe, S. and Schwindt, P. D. D. and Kitching, J.},
  journal={Nat. Photonics},
  volume={1},
  pages={649},
  year={2007},
  doi={10.1038/nphoton.2007.201}
}
@article{Forbes/Nat:2021,
author={Forbes, Andrew
and de Oliveira, Michael
and Dennis, Mark R.},
title={Structured light},
journal={Nat. Photonics},
year={2021},
month={Apr},
day={01},
volume={15},
number={4},
pages={253-262},
abstract={},
issn={1749-4893},
doi={10.1038/s41566-021-00780-4},
url={https://doi.org/10.1038/s41566-021-00780-4}
}

@article{Rubinsztein-Dunlop/JO:2017,
doi = {10.1088/2040-8978/19/1/013001},
url = {https://dx.doi.org/10.1088/2040-8978/19/1/013001},
year = {2017},
month = {nov},
publisher = {IOP Publishing},
volume = {19},
number = {1},
pages = {013001},
author = {Halina Rubinsztein-Dunlop and Andrew Forbes and M V Berry and M R Dennis and David L Andrews and Masud Mansuripur and Cornelia Denz and Christina Alpmann and Peter Banzer and Thomas Bauer and Ebrahim Karimi and Lorenzo Marrucci and Miles Padgett and Monika Ritsch-Marte and Natalia M Litchinitser and Nicholas P Bigelow and C Rosales-Guzmán and A Belmonte and J P Torres and Tyler W Neely and Mark Baker and Reuven Gordon and Alexander B Stilgoe and Jacquiline Romero and Andrew G White and Robert Fickler and Alan E Willner and Guodong Xie and Benjamin McMorran and Andrew M Weiner},
title = {Roadmap on structured light},
journal = {J. Opt.},
}

@article{Kopf/PRL:2025,
  title = {Conservation of Angular Momentum on a Single-Photon Level},
  author = {Kopf, L. and Barros, R. and Prabhakar, S. and Giese, E. and Fickler, R.},
  journal = {Phys. Rev. Lett.},
  volume = {134},
  issue = {20},
  pages = {203601},
  numpages = {7},
  year = {2025},
  month = {May},
  publisher = {American Physical Society},
  doi = {10.1103/PhysRevLett.134.203601},
  url = {https://link.aps.org/doi/10.1103/PhysRevLett.134.203601}
}

@article{Parigi/Nat:2015,
author={Parigi, Valentina
and D'Ambrosio, Vincenzo
and Arnold, Christophe
and Marrucci, Lorenzo
and Sciarrino, Fabio
and Laurat, Julien},
title={Storage and retrieval of vector beams of light in a multiple-degree-of-freedom quantum memory},
journal={Nat. Commun.},
year={2015},
month={Jul},
day={13},
volume={6},
number={1},
pages={7706},
abstract={},
issn={2041-1723},
doi={10.1038/ncomms8706},
url={https://doi.org/10.1038/ncomms8706}
}

@article{Pugatch/PRL:2007,
  title = {Topological Stability of Stored Optical Vortices},
  author = {Pugatch, R. and Shuker, M. and Firstenberg, O. and Ron, A. and Davidson, N.},
  journal = {Phys. Rev. Lett.},
  volume = {98},
  issue = {20},
  pages = {203601},
  numpages = {4},
  year = {2007},
  month = {May},
  publisher = {American Physical Society},
  doi = {10.1103/PhysRevLett.98.203601},
  url = {https://link.aps.org/doi/10.1103/PhysRevLett.98.203601}
}

@article{Richard/PRA:2024,
  title = {High-dimensional maximally entangled photon pairs in parametric down-conversion},
  author = {Bernecker, Richard and Baghdasaryan, Baghdasar and Fritzsche, Stephan},
  journal = {Phys. Rev. A},
  volume = {110},
  issue = {3},
  pages = {033718},
  numpages = {12},
  year = {2024},
  month = {Sep},
  publisher = {American Physical Society},
  doi = {10.1103/PhysRevA.110.033718},
  url = {https://link.aps.org/doi/10.1103/PhysRevA.110.033718}
}

@article{Schmiegelow/Nat:2016,
author={Schmiegelow, Christian T.
and Schulz, Jonas
and Kaufmann, Henning
and Ruster, Thomas
and Poschinger, Ulrich G.
and Schmidt-Kaler, Ferdinand},
title={Transfer of optical orbital angular momentum to a bound electron},
journal={Nat. Commun.},
year={2016},
month={Oct},
day={03},
volume={7},
number={1},
pages={12998},
abstract={},
issn={2041-1723},
doi={10.1038/ncomms12998},
url={https://doi.org/10.1038/ncomms12998}
}

@article{Lange/PRL:2022,
  title = {Excitation of an Electric Octupole Transition by Twisted Light},
  author = {Lange, R. and Huntemann, N. and Peshkov, A. A. and Surzhykov, A. and Peik, E.},
  journal = {Phys. Rev. Lett.},
  volume = {129},
  issue = {25},
  pages = {253901},
  numpages = {5},
  year = {2022},
  month = {Dec},
  publisher = {American Physical Society},
  doi = {10.1103/PhysRevLett.129.253901},
  url = {https://link.aps.org/doi/10.1103/PhysRevLett.129.253901}
}

@book{Gbur:2017,
  title={Singular optics},
  author={Gbur, Gregory J},
  year={2017},
  publisher={CRC press}
}

@article{Castellucci/PRL:2021,
  title = {Atomic Compass: Detecting 3D Magnetic Field Alignment with Vector Vortex Light},
  author = {Castellucci, Francesco and Clark, Thomas W. and Selyem, Adam and Wang, Jinwen and Franke-Arnold, Sonja},
  journal = {Phys. Rev. Lett.},
  volume = {127},
  issue = {23},
  pages = {233202},
  numpages = {6},
  year = {2021},
  month = {Nov},
  publisher = {American Physical Society},
  doi = {10.1103/PhysRevLett.127.233202},
  url = {https://link.aps.org/doi/10.1103/PhysRevLett.127.233202}
}

@article{Qiu/PR:2021,
author = {Shuwei Qiu and Jinwen Wang and Francesco Castellucci and Mingtao Cao and Shougang Zhang and Thomas W. Clark and Sonja Franke-Arnold and Hong Gao and Fuli Li},
journal = {Photon. Res.},
keywords = {CCD cameras; Cylindrical vector beams; Nitrogen vacancy centers; Optical fields; Tunable diode lasers; Vector beams},
number = {12},
pages = {2325--2331},
publisher = {Optica Publishing Group},
title = {Visualization of magnetic fields with cylindrical vector beams in a warm atomic vapor},
volume = {9},
month = {Dec},
year = {2021},
url = {https://opg.optica.org/prj/abstract.cfm?URI=prj-9-12-2325},
doi = {10.1364/PRJ.418522},
}

@article{Cai/LPR:24,
author = {Cai, Guoan and Tian, Ke and Wang, Zhaoying},
title = {Thermal Atomic Compass Based on Radially Polarized Beam},
journal = {Laser \& Photonics Rev},
volume = {18},
number = {11},
pages = {2400465},
keywords = {polarization selection absorption, radially polarized beam, thermal atomic compass},
doi = {https://doi.org/10.1002/lpor.202400465},
url = {https://onlinelibrary.wiley.com/doi/abs/10.1002/lpor.202400465},
abstract = {Abstract The relationship between the magnetic field direction and the spatial intensity distribution of a radially polarized light passing through a polarized thermal atom ensemble is investigated, which is intuitively presented in a polarization selection absorption effect of thermal atoms. The radially polarized light has a spatial axisymmetric polarization structure, which is set as the probe beam. If the direction of the applied magnetic field is transformed, the absorption of the alignment atomic system to special polarization components of the probe light is changed, resulting in a different absorption ratio. This allows the 3D vector direction of the magnetic field to be inferred by using only the absorption ratio and the projection coefficient of the transmission intensity pattern. Based on this, this work provides a compass based on a thermal atom system, demonstrating a new method for measuring the magnetic field direction in space.},
year = {2024}
}

@article{Ramakrishna/PRA:2024,
  title = {Interaction of vector light beams with atoms exposed to a time-dependent magnetic field},
  author = {Ramakrishna, S. and Schmidt, R. P. and Peshkov, A. A. and Franke-Arnold, S. and Surzhykov, A. and Fritzsche, S.},
  journal = {Phys. Rev. A},
  volume = {110},
  issue = {4},
  pages = {043101},
  numpages = {10},
  year = {2024},
  month = {Oct},
  publisher = {American Physical Society},
  doi = {10.1103/PhysRevA.110.043101},
  url = {https://link.aps.org/doi/10.1103/PhysRevA.110.043101}
}

@article{Ramakrishna/PRA:2025,
  title = {Interaction of a Poincar\'e beam with optically polarized atoms in the presence of a constant magnetic field},
  author = {Ramakrishna, S. and Fritzsche, S.},
  journal = {Phys. Rev. A},
  volume = {111},
  issue = {6},
  pages = {063110},
  numpages = {10},
  year = {2025},
  month = {Jun},
  publisher = {American Physical Society},
  doi = {10.1103/5r63-5hl1},
  url = {https://link.aps.org/doi/10.1103/5r63-5hl1}
}

@book{Johnson:2007,
  title={Atomic Structure Theory},
  author={Johnson, Walter R},
  year={2007},
  publisher={Springer, New York}
}

@article{Schulz/PRA:2020,
  title = {Generalized excitation of atomic multipole transitions by twisted light modes},
  author = {Schulz, S. A.-L. and Peshkov, A. A. and M\"uller, R. A. and Lange, R. and Huntemann, N. and Tamm, Chr. and Peik, E. and Surzhykov, A.},
  journal = {Phys. Rev. A},
  volume = {102},
  issue = {1},
  pages = {012812},
  numpages = {10},
  year = {2020},
  month = {Jul},
  publisher = {American Physical Society},
  doi = {10.1103/PhysRevA.102.012812},
  url = {https://link.aps.org/doi/10.1103/PhysRevA.102.012812}
}

@article{Matula/JPB:2013,
doi = {10.1088/0953-4075/46/20/205002},
url = {https://dx.doi.org/10.1088/0953-4075/46/20/205002},
year = {2013},
month = {oct},
publisher = {IOP Publishing},
volume = {46},
number = {20},
pages = {205002},
author = {O Matula and A G Hayrapetyan and V G Serbo and A Surzhykov and S Fritzsche},
title = {Atomic ionization of hydrogen-like ions by twisted photons: angular distribution of emitted electrons},
journal = {J. Phys. B},
}

@book{Blum:2012,
  title={Density Matrix Theory and Applications},
  author={Blum, Karl},
  year={2012},
  publisher={Springer, Berlin}
}

@book{Auzinsh:2010,
  title={Optically Polarized Atoms: Understanding Light-Atom Interactions},
  author={Auzinsh, Marcis and Budker, Dmitry and Rochester, Simon M},
  year={2010},
  publisher={Oxford University, Oxford}
}

@article{Wense:2020,
  title={The theory of direct laser excitation of nuclear transitions},
  author={von der Wense, Lars and Bilous, Pavlo V and Seiferle, Benedict and Stellmer, Simon and Weitenberg, Johannes and Thirolf, Peter G and P{\'a}lffy, Adriana and Kazakov, Georgy},
  journal={Eur. Phys. J. A},
  volume={56},
  pages={176},
  year={2020},
  publisher={Springer},
  url = {https://doi.org/10.1140/epja/s10050-020-00177-x}
}

@article{Tremblay/PRA:1990,
  title = {Optical pumping with two finite linewidth lasers},
  author = {Tremblay, P. and Jacques, C.},
  journal = {Phys. Rev. A},
  volume = {41},
  issue = {9},
  pages = {4989--4999},
  numpages = {0},
  year = {1990},
  month = {May},
  publisher = {American Physical Society},
  doi = {10.1103/PhysRevA.41.4989},
  url = {https://link.aps.org/doi/10.1103/PhysRevA.41.4989}
}

@article{Schmidt/PRA:2024,
  title = {Atomic photoexcitation as a tool for probing purity of twisted light modes},
  author = {Schmidt, R. P. and Ramakrishna, S. and Peshkov, A. A. and Huntemann, N. and Peik, E. and Fritzsche, S. and Surzhykov, A.},
  journal = {Phys. Rev. A},
  volume = {109},
  issue = {3},
  pages = {033103},
  numpages = {11},
  year = {2024},
  month = {Mar},
  publisher = {American Physical Society},
  doi = {10.1103/PhysRevA.109.033103},
  url = {https://link.aps.org/doi/10.1103/PhysRevA.109.033103}
}

@article{Fritzsche/CPC:2019,
  title={A fresh computational approach to atomic structures, processes and cascades},
  author={Stephan Fritzsche},
  journal={Comput. Phys. Commun.},
  volume={240},
  number={1},
  pages={1-14},
  year={2019},
  doi = {10.1016/j.cpc.2019.01.012},
}


@book{arfken2011/book,
  title={Mathematical methods for physicists: a comprehensive guide},
  author={Arfken, George B and Weber, Hans J and Harris, Frank E},
  year={2011},
  publisher={Academic press}
}


@book{Rose:1957,
  title={Elementary Theory of Angular Momentum},
  author={Rose, Morris Edgar},
  year={1957},
  publisher={John Wiley \& Sons, New York}
}

\preprint{}
\title{Unambiguous Vector Magnetometry with Structured Light in Atomic Vapor}

\author{S.~Ramakrishna}

\email[]{shreyas.ramakrishna@uni-jena.de}
\affiliation{Helmholtz-Institut Jena, D-07743 Jena, Germany}%
\affiliation{GSI Helmholtzzentrum f\"ur Schwerionenforschung GmbH, D-64291 Darmstadt, Germany}
\affiliation{Theoretisch-Physikalisches Institut, Friedrich-Schiller-Universit\"at Jena, D-07743 Jena, Germany}

\author{S.~Fritzsche}
\affiliation{Helmholtz-Institut Jena, D-07743 Jena, Germany}%
\affiliation{GSI Helmholtzzentrum f\"ur Schwerionenforschung GmbH, D-64291 Darmstadt, Germany}
\affiliation{Theoretisch-Physikalisches Institut, Friedrich-Schiller-Universit\"at Jena, D-07743 Jena, Germany}


\begin{abstract}

Absorption profiles of vector light upon interaction with atomic vapor carries distinct signatures of external magnetic field vector. However, this signature becomes ambiguous for anti parallel magnetic field vectors of equal magnitude, which makes their absorption profiles visually indistinguishable. To resolve this ambiguity, we present theoretical analysis of the interaction of vector light with optically polarized atoms immersed in reference and test magnetic fields. Furthermore, we demonstrate the complete characterization of the arbitrarily oriented test magnetic field via Fourier analysis of the absorption profile. This analysis reveals a one-to-one correspondence between the magnetic field properties and the profile’s contrast and rotational angle. Our findings open an avenue to design an optical vector atomic magnetometer based on structured light fields.
\end{abstract}

\newpage
\maketitle

\textit{Introduction.-}
Structured light fields typically exhibit inhomogeneous intensity and polarization profile across their beam cross section~\cite{Rubinsztein-Dunlop/JO:2017,Forbes/Nat:2021}. This unique feature have found numerous applications in the field of quantum entanglement~\cite{Kopf/PRL:2025,Richard/PRA:2024}, quantum memories~\cite{Parigi/Nat:2015,Pugatch/PRL:2007}, non-dipole atomic excitation~\cite{Schmiegelow/Nat:2016,Lange/PRL:2022} and many more. One can classify structured light modes into two categories: scalar modes, which have homogeneous polarization, and vector modes, which feature an inhomogeneous polarization texture~\cite{Gbur:2017}. As a prominent example, vector light fields exhibit an azimuthally varying linear state of polarization across the beam cross-section. This inherent in-homogeneity directly affects vector light’s interaction with an atomic target, causing the transition amplitude to be position-dependent. Consequently, light absorption by the atoms varies locally within the beam cross section. This phenomenon was recently  found to depend on the external magnetic field~\cite{Castellucci/PRL:2021}. Therefore, establishing a new atomic magnetometer scheme, allowing the field to be visualized via spatial absorption profile~\cite{Qiu/PR:2021,Cai/LPR:24,Ramakrishna/PRA:2024,Ramakrishna/PRA:2025}. This spatial-domain detection scheme provides a direct alternative to conventional atomic magnetometers~\cite{Shah/NP:2007,Budker:2013,Budker/N:2007}, which typically operate in the temporal domain using uniformly polarized light. 

For instance, in the presence of a transverse magnetic field, the absorption profile of the vector light acquires a flower-like pattern~\cite{Qiu/PR:2021}. However, a key limitation of this visual scheme is its ambiguity in determining the magnetic field's direction. For example, absorption profiles for magnetic fields of equal magnitude but opposite direction appear identical~\cite{Qiu/PR:2021,Cai/LPR:24}. Furthermore, existing works lacks a method to determine the strength of an arbitrarily oriented magnetic field—an essential capability for any vector atomic magnetometer.

 In this letter, we address the above mentioned two key limitations: we resolve the ambiguity between the anti-parallel magnetic field vectors and demonstrate a method for determination of the field’s strength of an arbitrarily oriented magnetic vector in space. Our theoretical model considers a vector beam probing an atomic vapor that is optically polarized and subjected to both a reference and a test magnetic field. Fourier analysis of the beam's absorption profile reveals a unique relationship, linking the field's properties directly to the profile's contrast and rotational angle. To illustrate these findings, we model the interaction of pump-probe fields with $^{87}$Rb atoms via the transition $5s \, ^2S_{1/2}$ ($F_{g}=1$) $\rightarrow$ $5p \, ^2P_{3/2}$ ($F_{e}=0$). These results open new avenues to design vector atomic magnetometers based on structured light fields.

\textit{Theory.-}
\begin{figure*}
    \centering
    \includegraphics[width=0.95\textwidth]{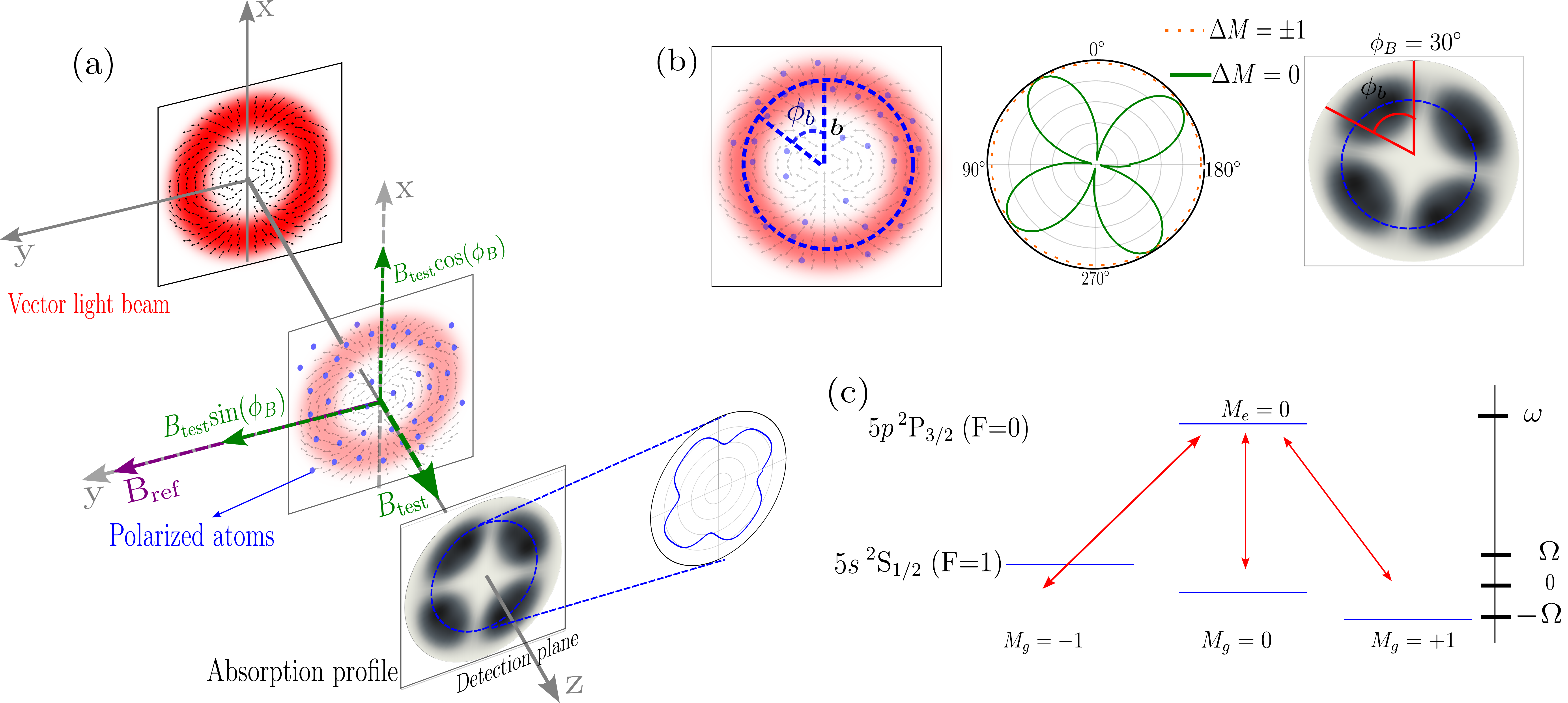}
    \caption{Geometrical setup of the system. (a) The vector light field of frequency $\omega$ with its inhomogeneous intensity and polarization texture propagates along the $z$ axis, and interacts with optically polarized atoms in the vapor cell. Here, the vapor cell is approximated by thin layer of randomly distributed atoms in the $xy$ plane. The reference field $\bm{B}_{\mathrm{ref}}$ is applied along the $y-$ axis, while the test $\bm{B}_{\mathrm{test}}$ magnetic field is a three dimensional vector. For such a system, the total magnetic field $\bm{B}_{T}$ lies suspended in the space and serves as the quantization axis (not shown in the diagram). The absorption profile of the vector light and its corresponding polar plot for a fixed radius of $b= 2000\, \mu m$ are shown in the detection plane. (b) The coupling between vector light and atomic target. The figure illustrates the interaction between vector light and polarized atoms for a $50$ mG test field applied at $\phi_{B} = 30^\circ$, and $0.1$ G reference magnetic field. The central polar plot shows the (normalized) transition amplitude for atoms along the blue dotted circle ($b = 2000\, \mu\text{m}$). This azimuthally varying interaction forms the absorption profile shown adjacent to it. Atoms in the dark regions absorb maximum light, as they are strongly coupled by the $M_{g} = 0 \rightarrow M_{e} = 0$ transition. Conversely, atoms in the lighter regions absorb less light, as they are mainly coupled via the $M_{g} = \pm 1 \rightarrow M_{e} = 0$ transition. (c) The energy level diagram of the rubidium atoms. In our scenario, the pump, linearly polarized plane wave couples $M_{g} = 0$ to $M_{e} = 0$ (weakly) and $M_{g} = \pm 1$ to $M_{e} = 0$ (strongly), leading to relatively maximum population in magnetic sublevels $M_{g} = 0$. }
    \label{fig:geometry}
\end{figure*}
We theoretically model a vapor cell with $^{87}$Rb atoms prepared in $F_{g} = 1$ hyperfine ground state at a temperature of $30\degree$ Celsius. The atoms are illuminated by pump and probe light fields that counter propagate along the $z$ axis, in the presence of magnetic fields. As shown in Fig.~\ref{fig:geometry} (a), we apply reference magnetic field along $y-$ axis $\bm{B}_{\mathrm{ref}} = B_{\mathrm{ref}}\, \bm{e}_y$ to measure the arbitrary three dimensional test field $\bm{B}_{\mathrm{test}} = B_{\mathrm{test}} \mathrm{cos}(\phi_{B})\,\bm{e}_x + B_{\mathrm{test}} \mathrm{sin}(\phi_{B})\, \bm{e}_y + B_{\mathrm{test}}\, \bm{e}_z$ in the space with an azimuthal angle of $\phi_{B}$. Therefore, the total magnetic field can be written as
\begin{equation}
    \bm{B}_{\mathrm{T}}(\phi_{B}) = \bm{B}_{\mathrm{ref}}\,+\,\bm{B}_{\mathrm{test}},
\end{equation}
which is the quantization axis of our system. Then, the atomic sublevels are split by a corresponding Larmor frequency $\Omega_{L}(\phi_{B}) = g_{F}\mu_{B}|\bm{B}_{\mathrm{T}}(\phi_{B})|/\hbar$, see  Fig.~\ref{fig:geometry} (c).  

Both the pump and probe light fields share the same angular frequency $\omega$, and beam width of $5$ mm. However, the pump light is assumed to be ten times more powerful than the probe. Upon interaction, these light fields drive a resonant transition between $5s \, ^2S_{1/2}$ ($F_{g}=1$) $\rightarrow$ $5p \, ^2P_{3/2}$ ($F_{e}=0$) levels in $^{87}$Rb atom. Our analysis considers a pump light field as a plane wave of linear polarization whose vector potential can be written as 
\begin{equation}
     \bm{A}^{\mathrm{(lin)}}(\bm{r}) = \frac{1}{\sqrt{2}} \left[ \bm{A}^{\mathrm{(circ)}}_{\lambda = +1}(\bm{r}) + \bm{A}^{\mathrm{(circ)}}_{\lambda = -1}(\bm{r}) \right],
\end{equation}
where $\bm{A}_{\lambda}^{(\mathrm{circ})} (\bm{r})$ is the vector potential of a circularly polarized plane wave with helicity $\lambda = \pm 1$. On the other hand, we consider the vector light in Bessel basis as the probe, and its vector potential can be written as
 \begin{align}
    \bm{A}^{\mathrm{(vec)}}(\bm{r}) =  \frac{1}{\sqrt{2}} \left[ \bm{A}^{\mathrm{(B)}}_{m_{\gamma} = -1, \, \lambda = +1}(\bm{r}) - \bm{A}^{\mathrm{(B)}}_{m_{\gamma} = +1, \, \lambda = -1}(\bm{r})  \right],
    \end{align}
where $\bm{A}^{\mathrm{(B)}}_{m_{\gamma}, \, \lambda}(\bm{r})$ denotes the vector potential of a circularly polarized Bessel beam in paraxial regime~\cite{Schulz/PRA:2020} with projection of total angular momentum $m_{\gamma} = |1|$, and opening angle $\theta_{k} = 0.008\degree$. The explicit expressions for vector potential of both plane wave and vector light is given in the Supplemental file. 

The interaction of these light fields with the atoms drives transitions between the ground $\left| \alpha_g F_g M_g \right>$ and excited $\left| \alpha_e F_e M_e \right>$ states. This process is described by the transition amplitude $V_{eg}$, calculated from the first-order matrix element~\cite{Johnson:2007}:
\begin{align}
V_{eg} =  \frac{1}{\alpha} \, \left\langle \alpha_{e}F_{e}M_{e} \left \vert \sum_{q} \bm{\alpha}_{q} \cdot \bm{A}(\bm{r}_{q})\right \vert \alpha_{g} F_{g} M_{g} \right \rangle, \label{eq:amp}
\end{align}
where $\alpha$ is the fine structure constant, $M$ is the projection of $F$ on the quantization axis, $\alpha_{g,e}$ are all other required quantum numbers, and $q$ sums over all atomic electrons and $\bm{\alpha}_{q}$ is the vector of Dirac matrices for the qth particle. The term $\bm{A}(\bm{r})$ represents the vector potential of the applied light field (either the plane wave or the vector beam). In our geometry, the linearly polarized plane wave (pump) couples all the three magnetic sublevels in the ground state to the excited state, however with maximum probability towards $M_{g} = \pm 1\rightarrow M_{e} = 0$ transition in comparison to $M_{g} = 0\rightarrow M_{e} = 0$ with the corresponding transition amplitude $V_{eg}^{\mathrm{(lin)}}$, as dictated by the selection rules. Thus, the pump light redistributes the ground state atomic population by accumulating relatively maximum population in the ground sublevels $M_{g} = 0$. 
In contrast, the vector beam's amplitude $V_{eg}^{\mathrm{(vec)}}(\bm{b})$ (explicit expressions in Supplemental file) varies with the position of atom $\bm{b} = (b,\phi_{b},0)$ in the beam cross section, as shown in Fig.~\ref{fig:geometry} (b). This variation originates from the inhomogeneous intensity and polarization profile of the vector beam. Because of the pump light-atom interaction, the probe couples sublevel $M_{g} = 0$ to the $M_{e} = 0$ relatively stronger compared to $M_{g} = \pm 1$ to the $M_{e} = 0$. 

The dynamics of the atomic level population, resulting from the atom-field interaction, are governed by the Liouville-von Neumann equation~\cite{Blum:2012} (see Supplemental file for explicit expressions)
\begin{equation}
    \frac{d}{dt}\hat{\rho}(t) = -\frac{i}{\hbar}[\hat{H}(t),\hat{\rho}(t)] + \hat{R}(t).
\end{equation}   
Here, density operator $\hat{\rho}(t)$ represents the atomic state, $\hat{H}(t)$ refers to the total Hamiltonian of the system, encompassing the interaction between atom and external fields, and the relaxation processes are taken into account phenomenologically via $\hat{R}(t)$~\cite{Tremblay/PRA:1990,Schmidt/PRA:2024}. Specifically, the operator $\hat{R}(t)$ accounts for incoherent relaxation processes: spontaneous emission, atomic transit motion ($\gamma_{\mathrm{transit}}$), and atom-atom collisions~\cite{Auzinsh:2010}. While the latter two processes are temperature dependent, at an operational temperature of $30^\circ\text{C}$ and in the absence of buffer gas, the rate of atomic collisions is negligible compared to the transit time decay of polarized atoms. Consequently, to mitigate transit time relaxation, we employ pump and probe light fields with sufficiently large beam widths ($5\,\text{mm}$).  Since $\gamma_{\mathrm{transit}}$ scales inversely with the beam width, this geometry ensures that the transit decay rate is slow relative to the optical pumping rate, resulting in effective atomic polarization.

Furthermore, in order to address the Doppler broadening associated with the atomic motion in the vapor cell, we have considered counter-propagation pump and probe light fields. This geometry allows us to select those atoms with longitudinal velocity class $v_{z} = 0$, as these atoms interact resonantly with the counter-propagating light fields (zero detuning). We note here that the exclusion of finite detuning and Doppler averaging would not alter the symmetry, rotational behavior, or Fourier structure of the absorption profile that underpins the proposed vector magnetometry scheme.

\textit{Absorption profile.-}
We analyze the steady state absorption profile of the vector light, which manifests as a flower-like pattern with four dark lobes, see Fig.~\ref{fig:geometry} (a). These dark lobes represents regions of maximum absorption in the beam cross section. We base our theoretical study on the assumption that these high-absorption regions directly correspond to the maximum excited-state population $\rho_{ee}$~\cite{Ramakrishna/PRA:2024}. Therefore, we obtain the absorption profile by calculating $\rho_{ee}$ across the beam cross section of the vector light field.

To supplement our assumptions, we simplify the problem by considering only atoms on the bright ring of the vector beam. The interaction of these atoms with the vector light is shown by the corresponding absolute transition amplitude plots. Upon interaction with a linearly polarized pump field, the atomic ground level population is redistributed. As a result, the incoming probe light couples the two sets of transitions $\Delta M = \pm 1, 0$ unequally. Mainly, the probe light relatively favors the $\Delta M = 0$ transition over the $\Delta M = \pm 1$ transitions, which is reflected in the absorption profile. In particular, the $\Delta M = 0$ transition dictates the position of the dark lobes in the absorption profile. Thus, these regions in the beam cross section with maximum excited state population appear darkest to the detector, see Fig.~\ref{fig:geometry}.



\textit{Elimination of directional ambiguity.-}
\begin{figure*}
    \centering
    \includegraphics[width=0.92\textwidth]{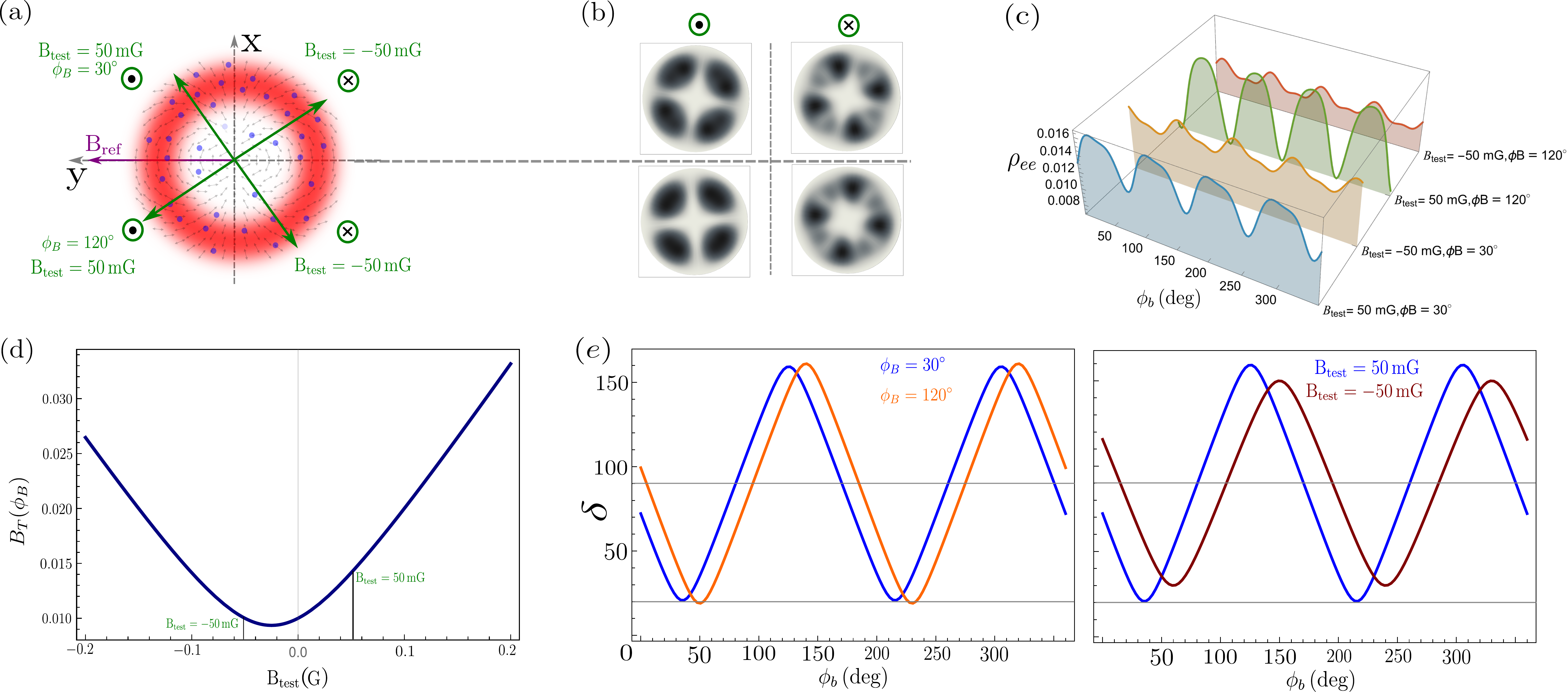}
    \caption{Absorption profiles of the vector beam. (a) The test magnetic field is shown at four different directions in a simplified geometry. (b) The corresponding four absorption profiles of the vector light (c) The corresponding population of excited state $\rho_{ee}$ for an atom at radial distance $b = 2000\, \mu m$ as a function of its azimuthal coordinate $\phi_{b}$. (d) The magnitude of $\bm{B}_{\mathrm{T}}$ as a function of test field's strength for $\phi_{B} = 120\degree$. (e) The angle between the local polarization vector and the quantization axis as a function of azimuthal coordinate of the atom. Left plot shows $\delta$ for two different azimuthal angle of test magnetic field of same strength, and right plot shows the variation for anti-parallel test magnetic field of same azimuthal angle $\phi_{B}$. For all these plots, the strength of reference $B_{\mathrm{ref}} = 0.1$ G} 
    \label{fig:quad}
\end{figure*}
We now analyze the absorption profiles for a scenario in which the test magnetic field has the same magnitude but opposite direction. In Fig.~\ref{fig:quad}, we display the vector light absorption profile for $50$ mG test magnetic field oriented along four different directions, together with reference field of $B_{\mathrm{ref}} = 0.1$ G. Among these four directions, the magnetic field vectors with $\phi_{B} = 30\degree$ and $120\degree$ are pointing out $\odot$ of the page with their anti parallel vectors pointing into the page $\otimes$, respectively, as shown in Fig.~\ref{fig:quad} (a). The absorption profiles corresponding to these anti-parallel vectors, as shown in Fig.~\ref{fig:quad} (b) are clearly distinguishable, a result not achieved in previous works~\cite{Qiu/PR:2021,Cai/LPR:24}. 
We observe that the absorption profile rotates as the azimuthal angle of the test magnetic field is varied from $\phi_{B} = 30\degree$ to $120\degree$. This is because the dark lobes form in the regions where the vector light strongly couples $M_{g} = 0 \rightarrow M_{e} = 0$. This particular transition will be maximum when the local polarization lies parallel to quantization axis, which can be denoted by $\delta$. In Fig.~\ref{fig:quad} (e) we show the variation of $\delta$ with $\phi_{b}$. As seen from this plot, the minimum value the $\delta$ takes coincides with the position of the dark region in the absorption profiles for $\phi_{B} = 30\degree$ and $120\degree$, in Fig.~\ref{fig:quad} (b). Furthermore, this minimum value of $\delta$ shifts as one varies the azimuthal angle of test magnetic field, see Fig.~\ref{fig:quad} (e) which results in the rotation of the absorption profile. Notably, the minimum angle between the local polarization and the quantization axis never reaches zero value completely due to the longitudinal component of the test magnetic field.

Moreover, in regions of the beam cross-section where the angle $\delta$ approaches $90^{\circ}$, the excited-state population decreases. Naturally, this condition favors transitions between the $M_{g} = \pm 1 \rightarrow M_{e} = 0$ sublevels. Since the pump light has not completely transferred the atomic ground-state population to $M_{g} = 0$, a small contribution to the excited-state population remains due to these transitions.

A comparison of the antiparallel fields ($\bm{B}_{\mathrm{test}} = \pm 50$ mG) reveals significant differences in the absorption profiles. As shown in Fig.~\ref{fig:quad} (e), switching from $50$ to $-50$ mG shifts the minimum of $\delta$, rotating the profile. Furthermore, the minimum $\delta$ is larger for $-50$ mG, indicating the quantization axis is less parallel to the polarization vector. Combined with the reduced total magnetic field strength for $-50$ mG [Fig.~\ref{fig:quad} (d)], these factors result in smaller dark lobes with lower contrast compared to the positive field case.

Thus, the combined effects of rotation and contrast ensure that each $\bm{B}_{\mathrm{test}}$ value produces a unique absorption profile. This ability to visually distinguish anti-parallel vectors stems from the non-zero reference magnetic field. Without this field, the system's response to $\bm{B}_{\mathrm{test}}$ and $-\bm{B}_{\mathrm{test}}$ is symmetric, leading to ambiguity in the generated absorption profile. The reference magnetic field breaks this symmetry, ensuring that total magnetic field have distinct values for each test magnetic field and in turn an unique absorption profile.

\textit{Detection of arbitrary magnetic field-} We next demonstrate how our scheme detects the magnitude of an arbitrarily directed test magnetic field. To do so, we quantify the unique relationship between the properties of test magnetic field and absorption profile through a Fourier analysis. In addition, we consider both, a test field with arbitrary strength and direction as well as a field whose longitudinal component is either significantly smaller or larger than the transverse components. Specifically, we consider two cases: (i) a weak longitudinal component, $\bm{B}_{\mathrm{test}} = B_{\mathrm{test}} \cos(\phi_{B})\,\bm{e}_x + B_{\mathrm{test}} \sin(\phi_{B})\, \bm{e}_y + 0.2\, B_{\text{test}}\, \bm{e}_{z}$; and (ii) a strong longitudinal component, $\bm{B}_{\mathrm{test}} = B_{\mathrm{test}} \cos(\phi_{B})\,\bm{e}_x + B_{\mathrm{test}} \sin(\phi_{B})\, \bm{e}_y + 5\, B_{\text{test}}\, \bm{e}_{z}$. For the latter case, we also examine the scenario in which the direction of the test magnetic field is reversed.

\begin{figure}
    \centering
    \includegraphics[width=0.5\textwidth]{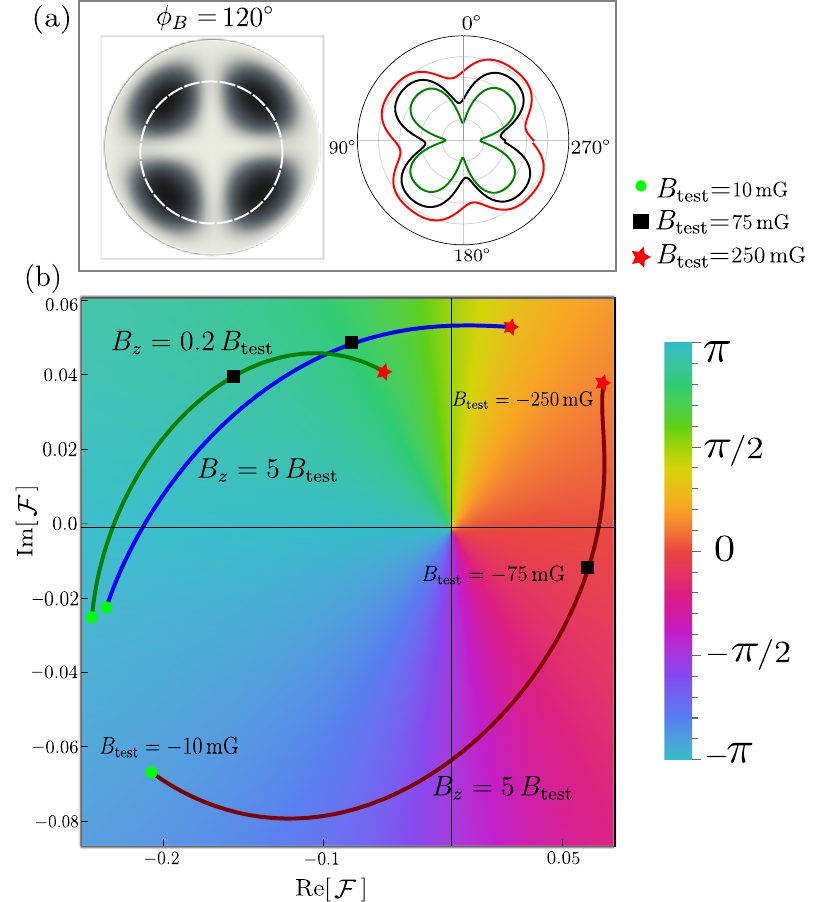}
    \caption{Variation of absorption profile with increasing strength of arbitrarily directed test magnetic field. (a) Left: Absorption profile of the vector light for a test magnetic field of azimuthal angle $\phi_{B} = 120\degree$ and strength of $10$ mG. Right: The corresponding polar plots for various strengths of constant test $B_{\mathrm{test}} = 10, 75$, and 250 mG considering the atoms to be positioned along the white dashed circle of radius $b = 2000\, \mu m$. (b)Trajectory of the normalized fourth Fourier harmonic, $\mathcal{F}_{4}$, for various strengths of the constant test magnetic field. Here, the strength of $B_{\mathrm{ref}} = 0.1$ G.}
    \label{fig:trajec}
\end{figure}

We first generate an equivalent polar plot of the absorption profile considering the atoms on a (white dashed) circle of radius $b = 2000\, \mu m$, see Fig.~\ref{fig:trajec} (a). For example, right figure in Fig.~\ref{fig:trajec} (a) illustrates polar plots for test magnetic field strengths $10$ (green), 75 (black), and 250 mG (red). For these plots, we set $B_{z} = 0.2\, B_{\text{test}}$, $\phi_{B} = 120\degree$ and reference magnetic field of $0.1$ G. To quantify these polar plots, we expand the expression of the population of excited state $\rho_{ee}$ in Fourier series.  Given the four-fold symmetry of the absorption profile, we extract the normalized fourth harmonic, $\mathcal{F}_{4} = \frac{a_4}{a_0}$, where $a_4$ and $a_0$ are the corresponding Fourier coefficients~\cite{arfken2011/book}. These normalized coefficients are then plotted in the complex plane to generate the trajectory plots, as shown in Fig.~\ref{fig:trajec} (b). These trajectories map how the absorption profile varies with an increase of test magnetic field strength. 

Each point on these trajectory curves is a complex number, $z = x+iy$, associated with the polar plot for a given $\bm{B}_{\mathrm{test}}$. The magnitude, $|z|$, quantifies the petal size (or contrast) of the four-fold pattern; a larger magnitude corresponds to more pronounced 'petals' and higher visibility of the lobes. The phase, $\arg(z) = \arctan(y/x)$, defines the physical orientation (rotation) of the profile, which is mapped to the background color scale in Fig.~\ref{fig:trajec} (b).

For $B_{z} = 0.2\, B_{\text{test}}$ and $\phi_{B} = 120^{\circ}$, the green solid line trajectory captures the variation of the absorption profile as the strength of the test field increases from $10$ to $250$ mG [see Fig.~\ref{fig:trajec}(b)]. Here, the magnitude of the complex number $|z|$ decreases steadily up to the second point ($150$ mG) and then rapidly thereafter. This can be visually confirmed in the polar plots, where the petal structure loses its prominent four-lobe shape as one goes from $10$ to $250$ mG. This behavior directly translates to the fact that the absorption profile loses its prominent four-lobe pattern as the test magnetic field strength is increased. Meanwhile, the phase increases from $-\pi$ approaching $+\pi/2$, indicating a counter-clockwise rotation of the petal structures.For the same azimuthal angle $\phi_{B} = 120^{\circ}$, if the longitudinal magnetic field component is larger than the transverse component ($5\,B_{\text{test}}$), then the blue solid line trajectory shows the variation. Since this curve lies closer to the center in comparison to the previous trajectory, the absorption profile's four-lobe pattern loses its significance at a faster rate. However, compared to the previous case, the blue trajectory sweeps a larger angle, which results in extensive rotation of the petal structure.    

As a next step, we show the trajectory for the case of an antiparallel test magnetic field. Once again, we choose $\phi_{B} = 120^{\circ}$ and $B_{z} = 5\, B_{\text{test}}$, but now the field strength is varied from $-10$ to $-250$ mG. The corresponding change in the absorption profile is shown by the maroon trajectory curve. In this case, the phase changes from approximately $-\pi$ all the way to $\pi/6$. This results in more extensive rotation of the petal structure in comparison to the other two cases. The obtained trajectory reveals that the absorption profiles for antiparallel magnetic fields are unique.       

\textit{Conclusions.-} We have theoretically demonstrated how vector light-matter interactions can be utilized to retrieve information regarding antiparallel magnetic field configurations. Unlike previous studies, our proposed scheme visually distinguishes the absorption profiles of antiparallel magnetic fields of equal magnitude. Furthermore, our approach enables the complete determination of an arbitrarily directed magnetic field vector directly from the spatial absorption profile of the vector beam.

We emphasize that the range of test magnetic field strengths was selected based on experimental feasibility~\cite{Qiu/PR:2021,Cai/LPR:24}. Furthermore, the primary objective of this work is to theoretically demonstrate the capability of unambiguously detecting an arbitrarily oriented magnetic field vector using structured light, rather than to optimize absolute sensitivity. Although the sensitivity can be qualitatively estimated using the metric $d\mathcal{F}_{4}/dB_{\mathrm{test}}$, a rigorous quantitative analysis would require incorporating noise sources and quantum-limited sensitivity. Such an analysis is beyond the scope of the present study and is reserved for future work. Furthermore, additional calculations confirm that our results are also valid for vector light with other orbital angular momentum ($m_{\ell}$) projections. For example, a vector light with $m_{\ell} = |1|$ gives rise to absorption profile with two-fold symmetry. Finally, these results advance our understanding of structured light-atom interaction and open new avenues for its application in optically pumped atomic magnetometers.


\textit{Acknowledgments}
We acknowledge support from the Research School of Advanced Photon Science of the Helmholtz Institute Jena.



\bibliography{Bibliography}

\end{document}